\documentclass{article}
\usepackage{graphicx}
\usepackage[english]{babel}
\usepackage[margin=1in]{geometry}
\usepackage{multicol}

\author{S. Chatterjee \footnote{Indian Insitute of Information Technology, Allahabad \texttt{$\lbrace$rs171@iiita.ac.in$\rbrace$}} \and B.S. Sanjeev \footnote{Indian Insitute of Information Technology, Allahabad \texttt{$\lbrace$sanjeev@iiita.ac.in$\rbrace$}} }
\title{
{\bfseries\Large Identification of Human Proteins vulnerable to multiple Organisms and their disease associations\bigskip}
}

\date{July 1, 2016}

\begin{document}
\maketitle

%

\tableofcontents


\begin{abstract}

While most studies emphasize on certain aspects of Pathogen-Host Interactions (PHI), such as the preferential attachment of bacteria or virus to its human receptor homolog, studies have attempted to methodically classify interactions among pathogenic proteins and their host proteins. Here we have analyzed 182 pathogens from The Pathogen-Host Interaction Search Tool (PHISTO)~\cite{TC13} and could identify the proteins/protein coding genes that act on both virus and bacteria. Importantly there were few proteins viz. P53 (Tumor protein p53), NFKB1 (Nuclear factor of kappa light polypeptide gene enhancer in B-cells 1), GBLP (Guanine nucleotide-binding protein subunit beta-2-like-1), TOX4 (TOX high mobility group box family member 4), PDIA1 (Protein disulfide-isomerase precursor), MHY9 (Myosin 9), RAC1 (Ras-related C3 botulinum toxin substrate 1), CCAR2 (Cell cycle and apoptosis regulator protein 2) and ILF3 (Interleukin enhancer binding factor 3) that were more susceptible to both bacterial and viral pathogens. Identification of such important interacting proteins (IIPs) can elicit significant insights for better understanding the molecular mechanisms of such pathogens that interact with the human host.


\end{abstract}

\section{Introduction}
\label{intro}

Pathogen-Host Interactions (PHI) are interactions that take place between pathogens (e.g. virus, bacteria, etc.) and their host (e.g. humans, plants). Pathogen-Host Interactions can be illustrated on a single-cell level (individual encounters of pathogen and host), on a molecular level (e.g. pathogenic protein binds to receptor on human cell), at the level of an organism (e.g. virus infects host), or on the population level (pathogen infections affecting a human population). Interactions among host and pathogen protein networks are called pathogen-host interactomes and investigating them may allow us to apprehend and be aware of the functioning of the host immune system and these important interacting proteins (IIPs) present in the host cell could be utilized as potential drug targets~\cite{LT15}. We have analyzed 3,905 interactions (edges) from 182 pathogens consisting of 1,188 proteins (nodes) that interact with 668 human proteins (nodes) from The Pathogen-Host Interaction Interaction Search Tool (PHISTO). The entries in PHISTO are curated and are also supported by strong experimental evidence as well as literature references.

\section{Materials and Methods}
\label{sec:1}
In graph theory and network analysis, we make use of several indicators of centrality to identify the most important vertices within a graph~\cite{DAD12},~\cite{ARS08}. Herein, for our undirected graph $\Gamma$, such that $\Gamma_p$ = ($V_p$, $E_p$), $V_p$ is the set of all nodes (proteins) of $H_p$ (Human) and $P_p$ (Pathogenic) and $E_p$ is the set of corresponding edges (interactions) wherein $E_p$ $\subseteq$ [$V_p]^2$. After analyzing the graph, we could pick out all the important interacting proteins (IIPs) in the human host that acted as hubs in the pathogen-host interactive graph and subsequently vulnerable to the pathogenic proteins. The pathogenic proteins were then meticulously short-listed and were linked to their respective pathogenic organisms/strains. We could list out 9 highly interacting hubs present in the human host that were involved with 786 pathogenic proteins. Herein, we have emphasized only on those human protein/ protein coding genes that were predominantly present in 5 or more viral pathogenic proteins and also subsequently present in 2 or more bacterial protein barring a few ones {\it viz.} RAC1, CCAR2 and ILF3.

\begin{figure}[htbp!]
\centering
\includegraphics[width=0.95\textwidth]{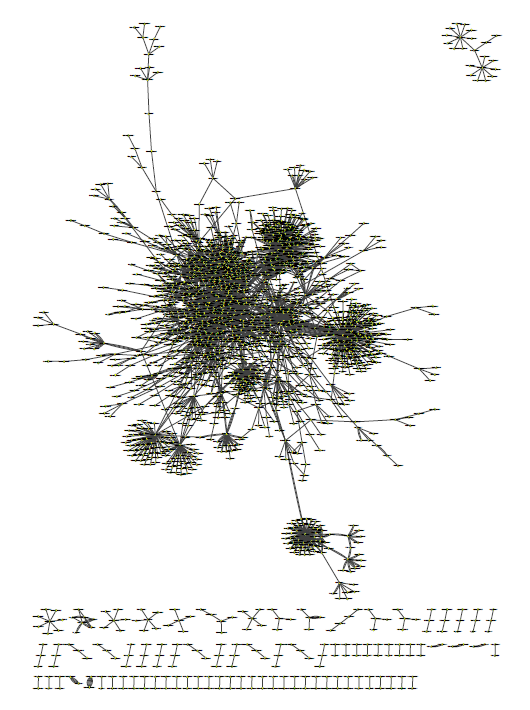}
\caption{The Pathogen-Host Network depicting the interactions (edges) between human and pathogenic proteins (nodes).}
\label{N_C_D}
\end{figure}

\begin{table}[htbp!]
\centering
\caption{Overview of PHI Networks}
\label{table:Analysis_of_PHI_Interactions}
\begin{tabular}{ll}
\hline\noalign{\smallskip}
Parameters	& Number (\#) \\
\noalign{\smallskip}\hline\noalign{\smallskip}
Total no. of pathogens 					& 182 	\\
No. of human proteins ($H_p$)			& 668	\\
No. of pathogen proteins ($P_p$)		& 1,188 \\ 
Total no. of proteins ($H_p$) + ($P_p$)	& 1,856 \\
Total no. of interactions 				& 3,905 \\
\noalign{\smallskip}\hline
\end{tabular}
\end{table}

\section{Results and Discussions}
\label{sec:2}

\section*{Network Analysis and Centrality Indices}
\label{sec:3}

\subsection{Node Degree Distribution}
\label{sec:5}
{\bf Definition:} The node degree distribution is an indicator of the number of nodes with a degree of {\it $k$}. The node degree of a particular node {\it $n$} is the number of interacting edges linked to {\it $n$}, in undirected networks.

\begin{figure}[htbp!]
\centering
\includegraphics[width=0.8\textwidth]{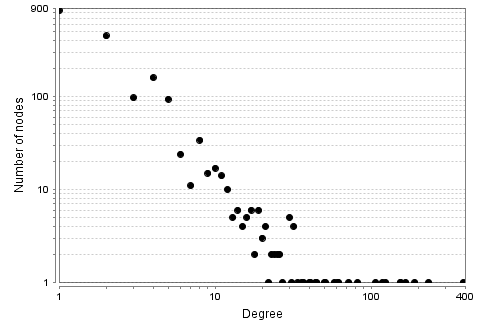}
\caption{Node Degree Distribution depicts the scale-free network topology of our pathogen-host network.}
\label{N_D_D}
\end{figure}

\subsection{Topological Coefficient}
\label{sec:6}
{\bf Definition:} Topological coefficient {\it $T(n)$} of a node {\it $n$} with {\it $k_n$} neighbors is defined as the number of neighbors shared between a pair of nodes, {\it $k_n$} and {\it $k_m$}, divided by the number of neighbors of node {\it $k_n$}:

\begin{equation}
T(n)={\frac {avg(J(n,m))}{k_n}}
\end{equation}

where the value of {\it $J(n,m)$} is the number of neighbors shared between the nodes {\it $n$} and {\it $m$}, plus one if there is a direct link between {\it $n$} and {\it $m$}. It gives a relative measure of a node that shares it's neighbors with other nodes and is defined for all nodes {\it $m$} that share at least one neighbor with {\it $n$}.

\begin{figure}[htbp!]
\centering
\includegraphics[width=0.8\textwidth]{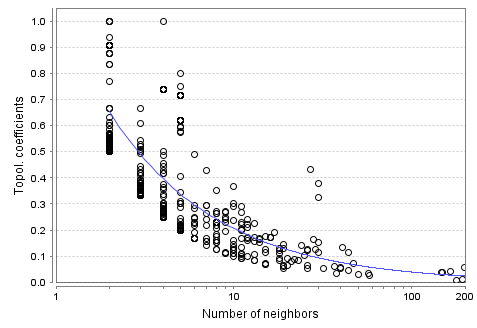}
\caption{Topological Coefficient is used to estimate the tendency of the nodes to have shared neighbors in the pathogen-host network.}
\label{T_C}
\end{figure}

\begin{table}[htbp!]
\centering
\caption{Summary of Network Statistics of PHI Networks}
\label{table:Network_Analysis_of_IIPs}
\begin{tabular}{ll}
\hline\noalign{\smallskip}
Parameter & Statistics	\\
\noalign{\smallskip}\hline\noalign{\smallskip}
Connected Components		& 82				\\
Network Diameter			& 17				\\
Network Centralization		& 0.105				\\
Shortest Paths				& 2,610,796 (75\%) 	\\ 
Characteristic Path Length	& 5.096				\\
Avg. number of Neighbors	& 3.357 			\\
Network Heterogeneity		& 3.231				\\
Multi-edge node pairs		& 496				\\
\noalign{\smallskip}\hline
\end{tabular}
\end{table}

\begin{table}[htbp!]
\centering
\caption{Overview of Important Interacting proteins (IIPs)}
\label{table:Analysis_of_IIPs}
\begin{tabular}{llll}
\hline\noalign{\smallskip}
Human Protein & Uniprot ID & Virus & Bacteria \\
\noalign{\smallskip}\hline\noalign{\smallskip}
P53		& P04637	&	35	&	5\\
NFKB1	& P19838	&	6	&	4\\
GBLP	& P63244	&	17	&	3\\
TOX4	& O94842	&	12	&	3\\
PDIA1	& P07237	&	9	&	2\\
MYH9	& P35579	&	6	&	3\\
RAC1	& P63000	&	3	&	7\\
CCAR2	& Q8N163	&   13	&	2\\
ILF3	& Q12906	&	23	&	1\\
\noalign{\smallskip}\hline
\end{tabular}
\end{table}

\begin{table}[htbp!]
\caption{Physiological significance of few Important Interacting Proteins (IIPs) and their Node Degree Distribution}
\label{table:Human_Protein_Functions}
\begin{tabular}{llll}
\hline\noalign{\smallskip}
S.No. & Protein & Degree & Biological Function and its significance \\
\noalign{\smallskip}\hline\noalign{\smallskip}
1 & P53 & 193 & Apoptotic process, Cell regulation, Cell cycle arrest, Chromatin assembly \\
2 & NFKB1 & 180 & Apoptotic process, Cell differentiation, Signal Transduction, Tumorigenesis \\
3 & GBLP & 40 & Assembly of signaling molecules, Cell cycle regulation, Promotes apoptosis \\
4 & TOX4 & 23 & Cell cycle progression, Control of Chromatin structure \\
5 & PDIA1 & 22 & Structural modification of exofacial proteins, Rearrangement of disulfide bonds \\
6 & MYH9 & 13 & Cytokinesis, Cytosketetal reorganization, Focal contacts formation \\
7 & RAC1 & 41 & Cell regulation, Cell polarization and growth-factor induced formation \\
8 & CCAR2 & 18 & Cellular integrity, Maintenance of genomic stability, Transcriptional activity \\
9 & ILF3 & 30 & Cell regulation, Regulates gene transcription \\
\noalign{\smallskip}\hline
\end{tabular}
\end{table}

\begin{table*}[htbp!]
\caption{List of few Important Interacting Proteins (IIPs) and their Disease Associations}
\label{table:Human_Disease_Gene_Associations}
\begin{tabular}{llll}
\hline\noalign{\smallskip}
S.No. & Associated Disease & List of IIPs associated with disease & Reference Database\\
\noalign{\smallskip}\hline\noalign{\smallskip}
1 & Adenocarcinoma & P53 / NFKB1 & CTD Human, CLINVAR \\
2 & Brain Ischemia & P53 / NFKB1 & CTD Human \\          
3 & Colonic Neoplasms & P53 / NFKB1 & CTD Human \\       
4 & Diabetes Mellitus & P53 / NFKB1 / RAC1 & CTD Human \\
5 & Kidney Failure & P53 / NFKB1 / MYH9 & CTD Human \\   
6 & Liver Carcinoma & P53 / RAC1 & CTD Human \\
7 & Liver Diseases & P53 / NFKB1 & CTD Human \\        
8 & Neoplastic Cell Transformation & P53 / NFKB1 / RAC1 & CTD Human \\
9 & Schizophrenia & P53 / NFKB1 & CTD Human, GWASCAT \\ 
\noalign{\smallskip}\hline
\end{tabular}
\end{table*}

\paragraph{}

The predominant proteins/ protein coding genes are described as follows:

The P53 (cellular tumor antigen or phosphoprotein 53) protein acts as a tumor suppressor protein in various tumor forms and it's main function is to induce growth arrest depending on the physiological circumstances~\cite{SK13}. Further, it also acts as a trans-activator involved cell-cycle regulation controlling a set of genes required for regulation~\cite{MM13},~\cite{SM98}.

As such, P53 has an important role in preserving stability of the genome by preventing the activity of mutagens in the genome~\cite{VM12},~\cite{RS99}.

The NFKB1 (NF-kappa-B) is a pleiotropic transcription factor involved in a number of signal transduction events that are instigated by stimulus that are related to cellular growth, cellular differentiation, inflammation, spread of tumorigenicity and ultimately cell death~\cite{BR04}.

GBLP (Guanine nucleotide-binding protein subunit beta-2-like 1) plays a role in the process of cell growth by prolonging the G0/G1 phase of the cell cycle~\cite{GB89} and is also involved in the conscription, assembly and/or regulation of a variety of signaling molecules. 

TOX4 (TOX high mobility group box family member 4) is a component of the PTW/PP1 phosphatase complex. It is involved in the control of chromatin structure and in the progression of cell cycle during the transition phase ~\cite{LU10}.

PDIA1 (Protein disulfide-isomerase) catalyzes the rearrangement of -S-S- bonds. The formation, breakage and rearrangement of -S-S- bonds is catalyzed by PDIA1 and hence may therefore cause structural modifications in exofacial proteins~\cite{BH11}.

MYH9 (Myosin 9) appears to play a role in maintaining the cell shape and reorganization of the cytoskeleton, cytokinesis and other specialized functions and plays an important role in focal contacts formation~\cite{B10}.

Interestingly, RAC1 (Ras-related C3 botulinum toxin substrate 1) is a GTPase associated with the plasma membrane and regulates the cellular responses by binding to a variety of effector proteins~\cite{NP11}. It was found to be more preferential to the bacterial proteins than the viral ones.

Interestingly, there are few proteins {\it viz.} CCAR2 and ILF3 which have more susceptibility to bind to the viral proteins than the bacterial ones. 

CCAR2 (Cell cycle and apoptosis regulator protein 2) is a multiprotein complex that acts at the interface between core mRNP particles and RNA polymerase II (RNAPII). It acts by inhibiting the deacetylase activity of SIRT1 leading to increasing acetylation levels of p53/TP53 and p53-mediated apoptosis~\cite{KC08}.

ILF3 (Interleukin enhancer-binding factor 3) may facilitate as a translation inhibitory protein which binds to mRNAs and other coding sequences of acid beta-glucosidase (GCase), inhibits binding to polysomes~\cite{KCB94} at the initiation phase of GCase mRNA translation.

\subsection{Association of IIPs with Disease Modules}
Diseases are not caused by a single gene, but rather in an orchestrated consequence of some abnormality that involves a cascade of interactions encompassing several cellular components (proteins/protein coding genes). We can well conclude that disease associated proteins are not confined to  local communities, but rather significantly incoherent.     

The result also suggest that while network analysis of such pathogen-host interactions can identify crucial local communities that point out important functionally related modules of the identified protein/protein coding genes, identification of a specific disease module can be regarded as a community detection problem. A community can be defined as a locally dense sub-graph in a network. Hence, identifying such sub-graphs can help us decode the distribution that a disease causes in particular.

\section{Summary and Conclusions}

We have identified the proteins/protein coding genes that were more susceptible to both bacterial and viral proteins. These are well targeted and act as potential drug targets. We have also inferred some disease associations of few proteins/ protein coding genes namely P53, MHY9, NKKB1 \& RAC1, that acted as hubs. The identified important interacting proteins (IIPs) also share common functionality {\it viz.} cellular differentiation, cell regulation, cell cycle arrest to cellular apoptosis. Our analysis of the pathogen-host interactions (PHIs) to identify the important interacting proteins (IIPs) that act on both bacterial and viral proteins can be practically applied over to other infectious fungal pathogens and protein-protein interactive networks (PPINs) as well.



\begin{thebibliography}{}
%
%

\bibitem{TC13}
S.~D. Tekir, T. \c{C}ak{\i}r, E. Ard{\i}\c{c}, A.~S. Say{\i}l{\i}rba\c{s}, G. Konuk, M. Konuk, H. Sar{\i}yer, A. U\u{g}urlu, \.{I}. Karadeniz, A. \"{O}zg\"{u}r, F.~E. Sevilgen, K.~\"{O}. \"{U}lgen, PHISTO: Pathogen-Host Interaction Search Tool, Bioinformatics., (Databases and ontologies):1357-1358, (2013).

\bibitem{LT15}
T. Lehnert, S. Timme, J. Pollm\"{a}cher, K. H\"{u}nniger, O. Kurzai, M.~T. Figge, Bottom-up modeling approach for the quantitative estimation of parameters in pathogen-host interactions, Front Microbiol 6, 608, (2015).

\bibitem{DAD12}
N.~T.Doncheva, Y. Assenov, F.~S. Domingues, M. Albrecht, Topological analysis and interactive visualization of biological networks and protein structures, Nature Protocols 7:670-685, (2012).

\bibitem{ARS08}
Y. Assenov, F. Ram\'{i}rez, S.~E. Schelhorn, T. Lengauer, M. Albrecht, Computing topological parameters of biological networks, Bioinformatics, 24(2):282-284, (2008).

\bibitem{SK13}
S. Surget, M.~P. Khoury, J.~C. Bourdon, Uncovering the role of p53 splice variants in human malignancy: a clinical perspective, OncoTargets and Therapy, 7: 57--68, (2013).

\bibitem{MM13}
T. Miki, T. Matsumoto, Z. Zhao, C.~C. Lee,
p53 regulates Period2 expression and the circadian clock, Nat. Commun., 4:2444-2444, (2013).

\bibitem{SM98}
E. Schneider, M. Montenarh, P. Wagner,
Regulation of CAK kinase activity by p53, Oncogene, 17: 2733-2741, (1998).

\bibitem{VM12}
A.~V. Vaseva, N.~D. Marchenko, K. Ji, S.~E. Tsirka, S. Holzmann, U.~M. Moll,
p53 opens the mitochondrial permeability transition pore to trigger necrosis, Cell, 149: 1536-1548, (2012).

\bibitem{RS99}
A.~P. Read and T. Strachan,
Human molecular genetics 2, Chapter 18: Cancer Genetics, New York: Wiley; ISBN 0-471-33061-2, (1999).

\bibitem{BR04}
S. Beinke, M.~J. Robinson, M. Hugunin, S.~C. Ley,
Lipopolysaccharide activation of the TPL-2/MEK/extracellular signal-regulated kinase mitogen-activated protein kinase cascade is regulated by IkappaB kinase-induced proteolysis of NF-kappaB1 p105, Mol. Cell. Biol., 24:9658-9667, (2004).

\bibitem{GB89}
F. Guillemot, A. Billault, C. Auffray, Physical linkage of a guanine nucleotide-binding protein-related gene to the chicken major histocompatibility complex, Proc. Natl. Acad. Sci. U.S.A., 86:4594-4598, (1989).

\bibitem{LU10}
J.~H. Lee, J. You, E. Dobrota, D.~G. Skalnik,
Identification and characterization of a novel human PP1 phosphatase complex, J. Biol. Chem., 285:24466-24476, (2010).

\bibitem{BH11}
S. Bi, P.~W. Hong, B. Lee, L.~G. Baum,
Galectin-9 binding to cell surface protein disulfide isomerase regulates the redox environment to enhance T-cell migration and HIV entry, Proc. Natl. Acad. Sci., 108:10650-10655, (2011).

\bibitem{B10}
V. Betapudi,
Myosin II motor proteins with different functions determine the fate of lamellipodia extension during cell spreading, PLoS ONE, 5:E8560-E8560, (2010).

\bibitem{NP11}
L. Naji, D. Pacholsky, P. Aspenstrom,
ARHGAP30 is a Wrch-1-interacting protein involved in actin dynamics and cell adhesion, Biochem. Biophys. Res. Commun., 409:96-102, (2011). 

\bibitem{KC08}
J.~E. Kim, J. Chen, Z. Lou,
DBC1 is a negative regulator of SIRT1, Nature, 451:583-586, (2008).

\bibitem{KCB94}
P.N. Kao, L. Chen, G. Brock, J. Ng, J. Kenny, A.~J. Smith, B. Corthesy,
Cloning and expression of cyclosporin A- and FK506-sensitive nuclear factor of activated T-cells: NF45 and NF90, J. Biol. Chem., 269:20691-20699, (1994).

\end{thebibliography}


\end{document}